\documentclass[showpacs,preprintnumbers,amsmath,amssymb]{revtex4}
\usepackage{graphicx}% Include figure files
\usepackage{subfigure}
\usepackage{dcolumn}% Align table columns on decimal point
\usepackage{bm}% bold math
\begin{document}
\title{Microscopic reversibility for classical open systems}
\author{Takaaki Monnai}%
\email{monnai@suou.waseda.jp}%
\affiliation{$*$Department of Applied Physics, Osaka City University,
3-3-138 Sugimoto, Sumiyoshi-ku, Osaka 558-8585, Japan}
\begin{abstract}
We rigorously show that the probability to have a specific trajectory of an externally perturbed classical open system satisfies a universal symmetry for Liouvillian reversible dynamics.  It connects the ratio between the probabilities of time forward and reversed trajectories to a degree of the time reversal asymmetry of the final phase space distribution. Indeed, if the final state is in equilibrium, then the forward and reversed net transition probabilities are equal, which gives a generalization of the detailed balance principle. 
On the other hand, when the external forcing maintains the system out of equilibrium, it expresses an asymmetry for the probabilities of the time forward and reversed trajectories. Especially, it gives a microscopic expression of the heat flowing to a system from a reservoir where the subdynamics seems like a Markovian stochastic process.

Also, it turns out that the expression of the microscopic reversibility holds both for the conservative and dissipative dynamics with an arbitrary initial state and external forcing.
\end{abstract}    
\pacs{05.70.Ln,05.40.-a}
\maketitle
\section{Introduction}
The time reversal invariance of the equations of motion equally allows both time forward and reversed processes.      
For the irreversible time evolution of macroscopic systems, however, we only observe one of them, the forward process. Indeed, the microscopic understanding of the macroscopic irreversibility is a long standing problem of the statistical physics.  
On the other hand, the reversed process can be observed through the fluctuation of quantities in mesoscopic systems\cite{Wang,Ciliberto}.     
Actually, the time reversal symmetry amounts to various fluctuation theorems and related nonequilibrium properties\cite{Evans1,Gallavotti,Jarzynski,Kurchan,Spohn,TasakiMatsui,Gelin,Jarzynski2,Crooks,Crooks2,Crooks3,Andrieux,Andrieux3,Andrieux2,Monnai,Monnai2,Utsumi,Esposito,Esposito2,Esposito3}. The fluctuation theorem concerns with the probability of the entropy production during the nonequilibrium process\cite{Gaspard}, and expresses a universal balance between the positive and negative entropy production within a model independent framework. Also the relation to the linear and nonlinear response theories has been established\cite{Spohn,Andrieux3,Andrieux2}, and thus it plays a fundamental role in the nonequilibrium statistical physics. 

For a Markovian stochastic dynamics, Ref.\cite{Crooks} provides a unified perspective which connects a fluctuation theorem to nonequilibrium work equality based on the so-called microscopic reversibility.
The validity of this theorem has been confirmed for the overdamped and underdamped Markovian Langevin dynamics\cite{Jarzynski3,Peliti} with the use of an expression of the heat essentially given in Ref.\cite{Sekimoto}.       

In this article, we rigorously investigate a corresponding microscopic reversibility for Liouvillian time evolutions for classical open systems.
We show that the microscopic reversibility connects the ratio between the probabilities of time forward and reversed trajectories to a degree of time reversal asymmetry of the final state distribution.  
We also explain some applications of the microscopic reversibility.  

This paper is organized as follows. In Sec.II, we prepare our model and explain the basic properties of Liouvillian dynamics. In Sec.III, microscopic reversibility is derived. Some applications are shown in Sec.IV.  Sec.V is devoted to a summary.  
\section{Model}     
We consider a deterministic dynamics in the phase space for a system interacting with reservoirs. 
Also our calculation accounts for the phase volume contraction, which does not necessarily contradict with the reversibility as explicitly shown in the appendix A.    
Let us denote the system- and reservoir- variables as $\Gamma_s$ and $\Gamma_r$.
They are composed of positions and momenta for Hamiltonian dynamics. 
The set of all variables $\{\Gamma_s,\Gamma_r\}$ is abbreviated as $\Gamma$.
An external perturbation acts on the system for $0\leq t\leq T$ in a certain forcing protocol. 
The initial state of the total system is arbitrary.
Note that a function of phase variables $A(\Gamma(t))$ evolves due to the temporal change of $\Gamma(t)$ 
\begin{eqnarray}
\dot{A}(\Gamma(t))&=&\frac{\partial A(\Gamma(t))}{\partial\Gamma(t)}\dot{\Gamma}(t) \nonumber \\
&=&L(t) A(\Gamma(t)),\label{observable}
\end{eqnarray} 
where $L(t)$ is the Liouvillian operator.
For Hamiltonian dynamics, it gives the Poisson bracket $L(t)A(\Gamma)= \{A(\Gamma),H(\Gamma,t)\}$ with a Hamiltonian $H(\Gamma,t)$.
On the other hand, the phase space distribution $\rho(\Gamma,t)$ evolves as 
\begin{eqnarray}
&&\frac{\partial \rho(\Gamma,t)}{\partial t} \nonumber \\
&=&-\frac{\partial}{\partial \Gamma_s}(\dot{\Gamma}_s\rho(\Gamma,t))-\frac{\partial}{\partial \Gamma_r}(\dot{\Gamma}_r\rho(\Gamma,t)) \nonumber \\
&=&-\left(\frac{\partial\dot{\Gamma}}{\partial\Gamma}+L(t)\right)\rho(\Gamma,t).\label{distribution}
\end{eqnarray}
Note that the sign of the Liouvillian operator is different between Eq.(\ref{observable}) and (\ref{distribution}). 
Then the conservation of probability is immediately obtained from
\begin{eqnarray}
\frac{d}{dt}\rho(\Gamma(t),t)&=&L(t)\rho(\Gamma(t),t)-\left(\frac{\partial\dot{\Gamma}(t)}{\partial\Gamma(t)}+L(t)\right)\rho(\Gamma(t),t) \nonumber \\
&=&-\frac{\partial\dot{\Gamma}(t)}{\partial\Gamma(t)}\rho(\Gamma(t),t)
\end{eqnarray} 
as 
\begin{equation}
\rho(\Gamma(t),t)=e^{-\int_0^t\frac{\partial\dot{\Gamma}(s)}{\partial\Gamma(s)}ds}\rho(\Gamma(0),0).\label{probability}
\end{equation}
\section{Derivation of microscopic reversibility}
We abbreviate the time evolution operator during $0\leq t\leq j\Delta T$ as ${\cal V}_j$. 
The trajectory is specified by the state at $t=j\Delta t$ as $\{\Gamma_s(0),\Gamma_s(\Delta t),...,\Gamma_s(N\Delta t)\}$ with a time step $\Delta t=\frac{T}{N}$. 
The forward probability $P_F(\{\Gamma_s(0),\Gamma_s(\Delta t),...,\Gamma_s(N\Delta t)\})$ to have  such a trajectory is given as
\begin{eqnarray}
&&P_F(\{\Gamma_s(0),\Gamma_s(\Delta t),...,\Gamma_s(N\Delta t)\}) \nonumber \\
&\equiv&\int d\Gamma\rho(\Gamma,0)\prod_{j=0}^N\delta(\Gamma_s(j\Delta t)-({\cal V}_j\Gamma)_s).\label{trajectory}
\end{eqnarray}
Eq.(\ref{trajectory}) actually gives the net transition probability that the system state at $t=j\Delta t$ is $\Gamma_s(j\Delta t)$.
Indeed, it satisfies the normalization condition for the net transition probability
\begin{eqnarray}
&&\int P_F(\{\Gamma_s(0),\Gamma_s(\Delta t),...,\Gamma_s(N\Delta t)\}) d\Gamma_s(0)\cdot\cdot\cdot d\Gamma_s(N\Delta t) \nonumber \\
&=&\int d\Gamma\rho(\Gamma,0) \nonumber \\
&=&1.
\end{eqnarray}
%For $N=1$, this definition gives the joint transition probability between the initial and final states.

Next, we define the probability for the time reversed trajectory $\{\Theta\Gamma_s(N\Delta t),\Theta\Gamma_s((N-1)\Delta t),...,\Theta\Gamma_s(\Delta t),\Theta\Gamma_s(0)\}$ where $\Theta$ is the time reversal operator acting on the total system. Note that $\Theta$ maps the set of all the positions and momenta at time $t$ $(q(t),p(t)$ as $\Theta(q(t),p(t))=(q(T-t),-p(T-t))$, i.e. the reversal with respect to time $T$.  
If there exists a magnetic field, we also reverse it, since the reversal of electron currents  amounts to the reversal of the corresponding magnetic field\cite{Andrieux2}.  
The probability for the reversed trajectory is given as
\begin{eqnarray}
&&P_R(\{\Theta\Gamma_s(N\Delta t),\Theta\Gamma_s((N-1)\Delta t),...,\Theta\Gamma_s(\Delta t),\Theta\Gamma_s(0)\}) \nonumber \\
&\equiv&\int d\Gamma\rho(\Gamma,T)\prod_{j=0}^N\delta(\Theta\Gamma_s(j\Delta t)-(\Theta{\cal V}_j{\cal V}_N^{-1}\Theta\Gamma)_s).\label{reverse}
\end{eqnarray}
We clarify that $\Theta{\cal V}_j{\cal V}_N^{-1}\Theta$ actually gives the reversed evolution operator from $t=0$ to $t=(N-j)\Delta t$. 
For simplicity, we here show this for Hamiltonian dynamics. The argument for thermostatted dynamics is similarly explained.  
Indeed, the time reversal consists of the reversal of the state $\Gamma=(q(t),p(t))\rightarrow\Theta\Gamma=(q(T-t),-p(T-t))$ and the reversal of the external forcing protocol. Here $q(t)$ and $p(t)$ are the set of all the positions and momenta.
The evolution operator is expressed by the time ordered product as 
\begin{eqnarray}
&&{\cal V}_N{\cal V}_j^{-1}={\rm T}\{e^{\int_{j\Delta t}^{N\Delta t}L(t)dt}\} \nonumber \\
&=&\lim_{\Delta t\rightarrow 0}e^{L(N\Delta t)\Delta t}e^{L((N-1)\Delta t)}\cdot\cdot\cdot e^{L((j+1)\Delta t)}e^{L(j\Delta t)},\label{evolution}
\end{eqnarray}
where the finite time approximation becomes rigorous in the continuous limit $\Delta t\rightarrow 0$ and $N\Delta t\rightarrow T$.
Using the property $\Theta L(t)=-L(t)\Theta$ and Eq.(\ref{evolution}), we have 
\begin{eqnarray}
&&\Theta{\cal V}_j{\cal V}_N^{-1}\Theta \nonumber \\
&=&\lim_{\Delta t\rightarrow 0}e^{L(j\Delta t)}e^{L((j+1)\Delta t)}\cdot\cdot\cdot e^{L((N-1)\Delta t)}e^{L(N\Delta t)}. \label{reversal}
 \end{eqnarray}
Next, we consider the time evolution of the reversed state $(\tilde{q}(t),\tilde{p}(t))\equiv(q(T-t),-p(T-t))$.
For this purpose, let us evaluate the value of an arbitrary function $A(\tilde{q},\tilde{p})$ at a time $t+dt$,
\begin{eqnarray}
&&A(\tilde{q}(t+dt),\tilde{p}(t+dt)) \nonumber \\
&=&A(\tilde{q}(t),\tilde{p}(t))+L(T-t)A(\tilde{q}(t),\tilde{p}(t))dt+O(dt^2) \nonumber \\
&=&A(q(T-t),-p(T-t))-\left(\frac{\partial A(q(T-t),-p(T-t))}{\partial q(T-t)}\frac{p(T-t)}{m}+\frac{\partial A(q(T-t),-p(T-t))}{\partial p(T-t)}\dot{p}(T-t)\right) dt+O(dt^2) \nonumber \\
&=&e^{L(T-t)dt}A(\tilde{q}(t),\tilde{p}(t)). \nonumber \\
\end{eqnarray}
In the first equality, we used that $(\tilde{q}(t),\tilde{p}(t))$ is also a solution of the equations of motion.
In the second equality, the sign of the terms of order $dt$ are different from that of the forward process, because of the reversal of momenta and the derivative of $A(q(T-t),-p(T-t))$ with respect to $p(T-t)$. 
Then one has
\begin{eqnarray}
&&\lim_{\Delta t\rightarrow 0}e^{L(j\Delta t)}e^{L((j+1)\Delta t)}\cdot\cdot\cdot e^{L((N-1)\Delta t)}e^{L(N\Delta t)}A(\tilde{q},\tilde{p})|_{(\tilde{q},\tilde{p})=(\tilde{q}(0),\tilde{p}(0))} \nonumber \\
&=&A(\tilde{q}(j\Delta t),\tilde{p}(j\Delta t)),
\end{eqnarray}
which shows that Eq.(\ref{reversal}) actually gives the time reversed evolution operator during $0\leq t\leq (N-j)\Delta t$.     
  
On the other hand, the forward transition probability is rewritten as         
\begin{eqnarray}
&&P_F(\{\Gamma_s(0),\Gamma_s(\Delta t),...,\Gamma_s((N-1)\Delta t),\Gamma_s(N\Delta t)\}) \nonumber \\
&=&\int d\Gamma\rho(\Gamma,0)\prod_{j=0}^N\delta(\Gamma_s(j\Delta t)-({\cal V}_j\Gamma)_s)\nonumber \\
&=&\int d({\cal V}_N\Gamma)|\frac{\partial\Gamma}{\partial({\cal V}_N\Gamma})|\rho({\cal V}_N^{-1}{\cal V}_N\Gamma,0)\prod_{j=0}^N\delta(\Gamma_s(j\Delta t)-({\cal V}_j{\cal V}_N^{-1}{\cal V}_N\Gamma)_s) \nonumber \\
&=&\int d({\cal V}_N\Gamma)e^{-\int_0^T\frac{\partial\dot{\Gamma}(t)}{\partial\Gamma(t)}dt}\rho({\cal V}_N^{-1}\Gamma',0) \nonumber \\
&&\times\prod_{j=0}^N\delta(\Gamma_s(j\Delta t)-({\cal V}_j{\cal V}_N^{-1}\Gamma')_s) \nonumber \\
&=&\int d\Gamma'\rho(\Gamma',T)\prod_{j=0}^N\delta(\Gamma_s(j\Delta t)-({\cal V}_j{\cal V}_N^{-1} \Gamma')_s) \nonumber \\
&=&\int d\Gamma'\rho(\Gamma',T)\prod_{j=0}^N\delta(\Theta\Gamma_s(j\Delta t)-(\Theta{\cal V}_j{\cal V}_N^{-1}\Gamma')_s) \nonumber \\
&=&\int d(\Theta\Gamma')\rho(\Theta\Gamma',T)\prod_{j=0}^N\delta(\Theta\Gamma_s(j\Delta t)-(\Theta{\cal V}_j{\cal V}_N^{-1}\Theta\Gamma')_s), \label{forward} \nonumber \\
\end{eqnarray}
where in the third equality, we changed the variable from $\Gamma$ to $\Gamma'={\cal V}_N\Gamma$, and applied a generalized Liouville theorem for $\Gamma'$, $\bigl|\frac{\partial\Gamma'}{\partial\Gamma}\bigr|=e^{\int_0^T\frac{\partial\dot{\Gamma}(t)}{\partial\Gamma(t)}dt}$.
In the fourth equality, we used the conservation of the probability Eq.(\ref{probability})  
\begin{equation}
e^{-\int_0^T\frac{\partial\dot{\Gamma}(t)}{\partial\Gamma(t)}dt}\rho({\cal V}_N^{-1}\Gamma',0)=\rho(\Gamma',T). \label{thermalization}
\end{equation}
Indeed the Jacobian for a change of variables $\Gamma\rightarrow{\cal V}_N\Gamma$ turns to be 
\begin{equation}
\bigl|\frac{\partial({\cal V}_N\Gamma)}{\partial\Gamma}\bigr|=e^{\int_0^t\frac{\partial\dot{\Gamma}(s)}{\partial\Gamma(s)}ds}\label{Jacobian}
\end{equation}
in the Appendix B.
In the fifth equality, we changed the variable as $\Gamma'\rightarrow\Theta\Gamma'$. 
Note that $\bigl|\frac{\partial(\Theta\Gamma')}{\partial\Gamma'}\bigr|=1$.
In the Dirac delta, $\Gamma_s$ is a set of variables and the time evolution operator acts on it, however, the quantities such as $\Gamma_s(0)$ and $\Gamma_s(T)$ are constant.  
Here $\rho(\Gamma,T)$ is the distribution function at $t=T$.

Comparing Eq.(\ref{forward}) with Eq.(\ref{reverse}), they look similar except for the time reversal in the statistical weight $\rho(\Theta\Gamma,T)$.
Therefore, the forward probability is directly connected to the reversed probability as
\begin{eqnarray}
&&\frac{P_F(\{\Gamma_s(0),\Gamma_s(\Delta t),...,\Gamma_s(T-\Delta t),\Gamma_s(T)\})}{P_R(\{\Theta\Gamma_s(T),\Theta\Gamma_s(T-\Delta t),...,\Theta\Gamma_s(\Delta t),\Theta\Gamma_s(0)\})} \nonumber \\
&=&\int d\Gamma \frac{\rho(\Theta\Gamma,T)}{\rho(\Gamma,T)} \frac{\rho(\Gamma,T)\prod_{j=0}^N\delta(\Theta\Gamma_s(j\Delta t)-(\Theta{\cal V}_j{\cal V}_N^{-1}\Theta\Gamma)_s)}{\int d\Gamma \rho(\Gamma,T)\prod_{j=0}^N\delta(\Theta\Gamma_s(j\Delta t)-(\Theta{\cal V}_j{\cal V}_N^{-1}\Theta\Gamma)_s)} \nonumber \\
&=&\bigl\langle\frac{\rho(\Theta\Gamma,T)}{\rho(\Gamma,T)}\bigr\rangle_R,
\end{eqnarray}
where $\langle \rangle_R$ is the average over the normalized transition probability that the reversed trajectory of the system state $\{\Theta\Gamma_s(N\Delta t),\Theta\Gamma_s((N-1)\Delta t),...,\Theta\Gamma_s(\Delta t),\Theta\Gamma_s(0)\}$ is observed.
Taking the continuous limit $\Delta t\rightarrow 0$ and $N\Delta t=T$, we finally obtain a model-independent symmetry of the net transition probability functionals 
\begin{equation}
\frac{P_F[\Gamma_s(t)]}{P_R[\Theta\Gamma_s(T-t)]}=\bigl\langle\frac{\rho(\Theta\Gamma,T)}{\rho(\Gamma,T)}\bigr\rangle_R, \label{microscopic}
\end{equation}        
for a trajectory $\Gamma_s(t)$ and its time reversal $\Theta\Gamma_s(T-t)$.   
Eq.(\ref{microscopic}) is a universal expression of the microscopic reversibility.
A notable feature of Eq.(\ref{microscopic}) is that it connects the ratio between the forward and reversed net transition probability functionals $P_F[\Gamma_s(t)]$ and $P_R[\Theta\Gamma_s(T-t)]$ to a degree of the time reversal asymmetry of the final state distribution. It expresses how difficult to achieve the time reversed trajectory compared with the forward trajectory. The physical implications of Eq.(\ref{microscopic}) are given in Sec.IV.
\section{Applications} 
If the system is submitted to a thermalization process, the final state would be close to thermal equilibrium which is invariant under the time reversal $\rho(\Theta\Gamma,T)=\rho(\Gamma,T)$.
In this case, the net transition probabilities of the forward and reversed trajectories are equal $P_F[\Gamma_s(t)]=P_R[\Gamma_s(T-t)]$, which is derived for quantum open systems submitted to a thermalization process\cite{Monnai3,Kawa}.
Especially, when the equilibrium state is maintained including the initial state, Eq.(\ref{microscopic}) corresponds to the detailed balance principle.

When there is only one large reservoir at an inverse temperature $\beta$ and the time evolution of the subsystem is well-described by a Markovian stochastic dynamics, Refs.\cite{Crooks,Peliti,Jarzynski3} show that the ratio of the conditional transition probabilities is equal to the exponential of the heat $e^{-\beta Q[\Gamma(t)]}$.
In such a case, Eq.(\ref{microscopic}) provides a microscopic expression of the heat $Q[\Gamma(t)]$. 
To derive the expression of the heat, we introduce a conditional probability.
The joint transition probability is rewritten as 
\begin{eqnarray}
&&P_F(\{\Gamma_s(0),\Gamma_s(\Delta t),....,\Gamma_s((N-1)\Delta t),\Gamma_s(N\Delta t)\}) \nonumber \\
&=&\rho(\Gamma_s(0),0)\int d\Gamma \frac{\rho(\Gamma_s(0),\Gamma_r,0)}{\rho(\Gamma_s(0),0)}\prod_{j=1}^N\delta(\Gamma_s(j\Delta t)-({\cal V}_j\Gamma)_s) \nonumber \\
&\equiv&\rho(\Gamma_s(0),0)P_F(\{\Gamma_s(0),\Gamma_s(\Delta t),....,\Gamma_s((N-1)\Delta t),\Gamma_s(N\Delta t)\}:\Gamma_s(0)),
\end{eqnarray}
where we defined a marginal distribution $\rho(\Gamma_s(0),0)\equiv\int d\Gamma_r\rho(\Gamma_s(0),\Gamma_r,0)$ and the conditional transition probability 
$P_F(\{\Gamma_s(0),\Gamma_s(\Delta t),....,\Gamma_s((N-1)\Delta t),\Gamma_s(N\Delta t)\}:\Gamma_s(0))$ with which we have a trajectory $\{\Gamma_s(0),\Gamma_s(\Delta t),....,\Gamma_s((N-1)\Delta t),\Gamma_s(N\Delta t)\}$ provided that the initial state of the system is $\Gamma_s(0)$.
The time reversed conditional probability is defined as well 
\begin{eqnarray}
&&P_R(\{\Theta\Gamma_s(T),\Theta\Gamma_s(T-\Delta t),...,\Theta\Gamma_s(\Delta t),\Theta\Gamma_s(0)\}:\Theta\Gamma_s(T)) \nonumber \\
&\equiv&\int d\Gamma\frac{\rho(\Theta\Gamma_s(T),\Theta\Gamma_r,T)}{\rho(\Theta\Gamma_s(T),T)}\prod_{j=0}^N\delta(\Theta\Gamma_s(j\Delta t)-(\Theta{\cal V}_j{\cal V}_N^{-1}\Theta\Gamma)_s)
\end{eqnarray}
 with a marginal distribution $\rho(\Gamma_s,T)\equiv\int d\Gamma_r\rho(\Gamma_s,\Gamma_r,T)$. 

With the use of conditional probabilities, the microscopic reversibility is expressed as
\begin{equation}
\frac{P_F[\Gamma_s(t):\Gamma(0)]}{P_R[\Theta\Gamma_s(T-t):\Theta\Gamma_s(T)]}=\frac{\rho(\Theta\Gamma_s(T),T)}{\rho(\Gamma_s(0),0)}\bigl\langle\frac{\rho(\Theta\Gamma,T)}{\rho(\Gamma,T)}\bigr\rangle_R.
\end{equation}
On the other hand, this quantity is equal to $e^{-\beta Q[\Gamma(t)]}$\cite{Crooks,Peliti,Jarzynski3}.
Thus we obtain a microscopic expression of the heat
\begin{equation}
Q[\Gamma(t)]=-\frac{1}{\beta}\log\frac{\rho(\Theta\Gamma_s(T),T)}{\rho(\Gamma_s(0),0)}\bigl\langle\frac{\rho(\Theta\Gamma,T)}{\rho(\Gamma,T)}\bigr\rangle_R. \label{dissipation}
\end{equation}
This expression gives an important condition Eq.(\ref{quasi}) for thermalization processes.
Let us consider the following case.
The system is initially in equilibrium described by a canonical ensemble $\frac{1}{Z_s(0)}e^{-\beta H_s(\Gamma_s,0)}$ with the system Hamiltonian $H_s(\Gamma_s,0)$, and externally perturbed during some transient time.
Then the total system is left untouched so that it would thermalize.
At the final time $T$, the total system is in equilibrium again, and therefore $\bigl\langle\frac{\rho(\Theta\Gamma,T)}{\rho(\Gamma,T)}\bigr\rangle_R=1$.
Especially, the subsystem is described by another canonical ensemble $\frac{1}{Z_s(T)}e^{-\beta H_s(\Gamma_s,T)}$.
Then the heat averaged over the trajectories $Q[\Gamma(t)]$ is calculated from Eq.(\ref{dissipation}) and canonical ensembles at $t=0,T$ as
\begin{eqnarray}
&&\beta\langle Q[\Gamma(t)]\rangle \nonumber \\
&=&\beta\left(\langle H_s(\Gamma_s(T),T)\rangle-\langle H_s(\Gamma_s(0),0)\rangle-\Delta F_s\right) \nonumber \\
&=&-\frac{\Delta S}{k_B}, \label{quasi}
\end{eqnarray}
which shows that equality holds in the Clausius inequality for the entropy change $\Delta S$. 
In the first equality, we identified the averaged difference of the Hamiltonians as an internal energy change $\Delta E$. Here the free energy difference $\Delta F_s=-\frac{1}{\beta}\log\frac{Z_s(T)}{Z_s(0)}$ is calculated as $\Delta F_s=\Delta E-\frac{\Delta S}{\beta k_B}$. 
The appearance of a reasonable condition Eq.(\ref{quasi}) for static cases confirms the validity of the expression of the heat Eq.(\ref{dissipation}).
\section{Summary}
In conclusion, an expression of the microscopic reversibility for classical open systems is rigorously derived by taking into account the conservation of the probability. Remarkably, it directly connects the probabilities for time forward and reversed trajectories as an time reversal asymmetry of the final state distribution. The probability of the forward and reversed trajectory would be different in the presence of dissipation, which is expressed by the asymmetry factor.  Also, the initial state and interaction between the system and reservoir are arbitrary.  
Then its properties are investigated.
If the total system would thermalize after the initial transient\cite{Monnai3,Monnai4,Vandenbroeck,
Lebowitz,Sugita,Reimann,Rigol}, the forward and reversed probabilities are almost equal, which is consistent with the results for qunatum systems based on the two-point measurement\cite{Monnai3,Kawa}. This would be regarded as a nonequilibrium generalization of the detailed balance principle. 
We also derived a microscopic expression of the heat, which reasonably yields Clausius equality for quasi static processes.    
\section{Acknowledgment}
This work is financially supported by JSPS Research Fellowship under the Grant 22$\cdot$7744.
\appendix
\section{Conservative and dissipative dynamics}
Here, we show two concrete examples of both conservative and dissipative dynamics where the microscopic reversibility holds. This would make clear how the conservation of the probability actually looks like.   
Discussions on the conservative case is especially useful to compare with the corresponding quantum version\cite{Monnai3}.   
The most fundamental conservative dynamics is the Hamiltonian dynamics.
Since the perturbation acts only on the system, the total Hamiltonian is given as
\begin{equation}
H=H_s(t)+H_r+H_{sr},
\end{equation}
where $H_s(t)$, $H_r$, and $H_{sr}$ are Hamiltonians of the system, reservoirs, and interactioon between them, respectively.
Unlike the quantum case where the system is submitted to a thermalization after a quench, the interaction energy $H_{sr}$ needs not be small compared with bulk energies $H_s$ and $H_r$.
Also the time dependence of $H_s(t)$ is arbitrary and $\rho(\Gamma,T)$ can be out of equilibrium.
These points are in marked contrast with the microscopic reversibility of quantum open systems.
In the present case, calculations are greatly simplified due to the conservative property $\frac{\partial\dot{\Gamma}}{\partial\Gamma}=0$, which amounts to
$|\frac{\partial({\cal V}\Gamma)}{\partial\Gamma}|=1$, and conservation of the probability is  $\rho({\cal V}^{-1}\Gamma,0)=\rho(\Gamma,T)$, which accounts for the Hamiltonian time evolution during $0\leq t\leq T$.  
    
As an important case of reversible but non-conservative system, we focus on Nose-Hoover dynamics, since it would not be obtained by coarse graining of Hamiltonian dynamics. Note that the dynamics is dissipative due to the presence of a thermostat. 
Analysis here provides a concrete example where the dissipation actually comes into the calcuation.   
Let us consider the $D$ dimensional $N$-particle system, which obeys the equation of motion
\begin{eqnarray}
&&\dot{x}_i=\frac{p_i}{m} \nonumber \\
&&\dot{p}_i=-\frac{\partial V(x_1,...,x_{DN},t)}{\partial x_i}-\zeta p_i \nonumber \\
&&\dot{\zeta}=\alpha\frac{\sum_{i=1}^{DN} (p_i^2-\langle p_i^2\rangle)}{\langle p_i^2\rangle},
\end{eqnarray}
where $m$ is the mass, $\zeta$ mimics the thermostat and $\alpha$ is a measure of the relaxation rate.
$\langle\frac{p_i^2}{m}\rangle=\frac{1}{\beta}$ gives the kinetic temperature.  
We regard $\Gamma_s=\{x_i,p_i\}$ as system variables.  
The Liouvillian operator acts on the distribution $\rho(\Gamma_s,\zeta,t)$ as
\begin{eqnarray}
&&\frac{\partial}{\partial t}\rho(\Gamma_s,\zeta,t) \nonumber \\
&=&-\sum_{i=1}^{DN}(\frac{p_i}{m}\frac{\partial}{\partial x_i}\rho \nonumber \\
&&-\frac{\partial}{\partial p_i}
(\frac{\partial V(\Gamma_s,\zeta,t)}{\partial x_i}+\zeta p_i)\rho+\frac{\partial}{\partial\zeta}\alpha\frac{p_i^2-\langle p_i^2\rangle}{\langle p_i^2\rangle}\rho) \nonumber \\
&=&-(L(t)-DN\zeta(t))\rho(\Gamma_s,\zeta,t).
\end{eqnarray}
Then the phase volume varies in time
\begin{equation}
\frac{d}{dt}\rho(\Gamma_s(t),\zeta(t),t)=DN\zeta(t)\rho(\Gamma_s(t),\zeta(t),t),
\end{equation}
and 
\begin{equation} 
\rho(\Gamma_s(t),\zeta(t),t)=e^{\int_0^t DN\zeta(s)ds}\rho(\Gamma_s(0),\zeta(0),0),
\end{equation}
which corresponds to the Liouville theorem. 
Let us abbreviate the time evolution operator of the variable $\Gamma_s$ and $\zeta$ as ${\cal V}$.
We also denote $({\cal V}\Gamma)_r$ as ${\cal V}\zeta$.
The conservation of the probability is then 
\begin{equation}
e^{\int_0^T DN\zeta(t)dt}\rho(({\cal V}^{-1}\Gamma)_s,{\cal V}^{-1}\zeta,0)=\rho(\Gamma,\zeta,T)
\end{equation}
and the Jacobian is explicitly calculated as 
\begin{equation}
\bigl|\frac{\partial(({\cal V}\Gamma)_s,{\cal V}\zeta)}{\partial(\Gamma_s,\zeta)}\bigr|=e^{-\int_0^T DN\zeta(s)ds}.
\end{equation} 
Despite the phase volume contraction, the microscopic reversibility holds as the conservative case.
\section{Generalized Liouville theorem}
In Eq.(\ref{forward}), the Jacobian  
\begin{equation}
\bigl|\frac{\partial({\cal V}_N\Gamma)}{\partial\Gamma}\bigr|=e^{\int_0^t\frac{\partial\dot{\Gamma}(s)}{\partial\Gamma(s)}ds}
\end{equation}
is used.
To show this, we divide the time interval $[0,t]$ into $NM$ many short intervals.
First $[0,t]$ is decomposed to $N$ intervals $[n\Delta t,(n+1)\Delta t]$ with $\Delta t=\frac{t}{N}$ so that external perturbation is approximately constant for each interval. 
Then the interval $[n\Delta t,(n+1)\Delta t]$ is further partitioned as $[n\Delta t+m\Delta s,n\Delta t+(m+1)\Delta s]$ with $\Delta s=\frac{\Delta t}{M}$.
We denote the $j$-th component of $\Gamma$ as $\Gamma_j$ and consider the case of two components for simplicity.
Up to the first order of $\Delta s$, we have $\Gamma_j(n\Delta t+(m+1)\Delta s)=\Gamma_j(n\Delta t+m\Delta s)+\dot{\Gamma}_j(n\Delta t+m\Delta s)\Delta s$.
And the Jacobian for the change of variables $\Gamma(n\Delta t+m\Delta s)\rightarrow\Gamma(n\Delta t+(m+1)\Delta s)$ is 
\begin{eqnarray}
&&\left|
\begin{array}{cc}
1+\frac{\partial\dot{\Gamma}_1(t_{nm})}{\partial\Gamma_1(t_{nm})}\Delta s & \frac{\partial\dot{\Gamma}_2(t_{nm})}{\partial\Gamma_1(t_{nm})}\Delta s \nonumber \\
\frac{\partial\dot{\Gamma}_1(t_{nm})}{\partial\Gamma_2(t_{nm})}\Delta s & 1+\frac{\partial\dot{\Gamma}_2(t_{nm})}{\partial\Gamma_2(t_{nm})}\Delta s
\end{array}
\right| \nonumber \\
&=&1+\sum_j\frac{\partial\dot{\Gamma}_j(t_{nm})}{\partial\Gamma_j(t_{nm})}\Delta s+O(\Delta s^2),
\end{eqnarray} 
where $t_{nm}=n\Delta t+m\Delta s$.
For a fixed $n$, the value of the determinant is regarded as constant up to the first order of $\Delta s$, and the Jacobian for the change of variables $\Gamma(t_{n0})\rightarrow\Gamma(t_{nM})$ is calculated as
\begin{equation}
\lim_{M\rightarrow\infty}\left(1+\sum_j\frac{\partial\dot{\Gamma}_j(n\Delta t)}{\partial\Gamma_j(n\Delta t)}\frac{\Delta t}{M}\right)^M=e^{\sum_j\frac{\partial\dot{\Gamma}_j(n\Delta t)}{\partial\Gamma_j(n\Delta t)}\Delta t}.
\end{equation}
Repeating a similar procedure for $n$ and multiplying each factor, Eq.(\ref{Jacobian}) is derived.

\end{document}